\documentclass[prd,twocolumn,showpacs,amsmath,amssymb,nofootinbib]{revtex4-1}

\usepackage{epsf}
\usepackage{graphicx}
\usepackage{epsfig}
\usepackage{xcolor}
\usepackage{subfigure}
\usepackage{pstricks}
\usepackage{pst-node}
\usepackage{rotating}
\usepackage{graphics}
\usepackage{latexsym}
\usepackage{dcolumn}
\usepackage{bm}
\usepackage{times}
\usepackage{indentfirst}
\usepackage{float}
\usepackage{overpic}
\usepackage{amsmath}
\usepackage{appendix}
\usepackage{upgreek}
\usepackage{chngcntr}
\usepackage{amssymb}%
\usepackage{pifont}%
\usepackage[top=0.5in,bottom=0.8in,left=0.9in,right=0.9in]{geometry}
\usepackage{subfigure}
\usepackage{lineno}

\usepackage{multirow}
\usepackage{makecell}

\usepackage{hyperref}
\hypersetup{
	colorlinks=true,
	linkcolor=blue,
	filecolor=blue,
	urlcolor=blue,
	citecolor=blue,
}

\newcommand{\llb}{\Lambda\bar{\Lambda}}

\newcommand{\pip}{\pi^+}
\newcommand{\pim}{\pi^-}

\newcommand{\pbar}{\bar{p}}

\newcommand{\EE}{e^+e^-}

\newcommand{\kkkk}{K^+K^-K^+K^-}

\newcommand{\jpsi}{J/\psi}
\newcommand{\ar}{\rightarrow}

		\newcommand{\bfg}{\begin{figure}}
			\newcommand{\efg}{\end{figure}}
		\newcommand{\bitm}{\begin{itemize}}
			\newcommand{\eitm}{\end{itemize}}
		\newcommand{\bnum}{\begin{enumerate}}
			\newcommand{\enum}{\end{enumerate}}
		\newcommand{\btbl}{\begin{table}}
			\newcommand{\etbl}{\end{table}}
		\newcommand{\btbu}{\begin{tabular}}
			\newcommand{\etbu}{\end{tabular}}
		\newcommand{\bcl}{\begin{center}}
			\newcommand{\ecl}{\end{center}}
		
		\newcommand{\beq}{\begin{equation}}
			\newcommand{\eeq}{\end{equation}}
		\newcommand{\beqr}{\begin{eqnarray}}
			\newcommand{\eeqr}{\end{eqnarray}}
\begin{document}
	
\title{\boldmath Measurement of the $\EE\ar\llb$ cross section from
  threshold to 3.00~GeV using events with initial-state radiation}
	
\author{
M.~Ablikim$^{1}$, M.~N.~Achasov$^{13,b}$, P.~Adlarson$^{73}$, R.~Aliberti$^{34}$, A.~Amoroso$^{72A,72C}$, M.~R.~An$^{38}$, Q.~An$^{69,56}$, Y.~Bai$^{55}$, O.~Bakina$^{35}$, I.~Balossino$^{29A}$
% ...
\\
\vspace{0.2cm}
(BESIII Collaboration)\\
\vspace{0.2cm} {\it
	$^{1}$ Institute of High Energy Physics, Beijing 100049, People's Republic of China\\
% ...
$^{13}$ G.I. Budker Institute of Nuclear Physics SB RAS (BINP), Novosibirsk 630090, Russia\\
% ...
$^{29}$ INFN Sezione di Ferrara, (A)INFN Sezione di Ferrara, I-44122, Ferrara, Italy; (B)University of Ferrara,  I-44122, Ferrara, Italy\\
% ...
$^{34}$ Johannes Gutenberg University of Mainz, Johann-Joachim-Becher-Weg 45, D-55099 Mainz, Germany\\
	$^{35}$ Joint Institute for Nuclear Research, 141980 Dubna, Moscow region, Russia\\
% ...
$^{38}$ Liaoning Normal University, Dalian 116029, People's Republic of China\\
% ...
$^{55}$ Southeast University, Nanjing 211100, People's Republic of China\\
	$^{56}$ State Key Laboratory of Particle Detection and Electronics, Beijing 100049, Hefei 230026, People's Republic of China\\
% ...
$^{69}$ University of Science and Technology of China, Hefei 230026, People's Republic of China\\
% ...
$^{72}$ University of Turin and INFN, (A)University of Turin, I-10125, Turin, Italy; (B)University of Eastern Piedmont, I-15121, Alessandria, Italy; (C)INFN, I-10125, Turin, Italy\\
	$^{73}$ Uppsala University, Box 516, SE-75120 Uppsala, Sweden\\
% ...
$^{b}$ Also at the Novosibirsk State University, Novosibirsk, 630090, Russia\\
% ...
}
}

\date{\today}

\renewcommand{\abstractname}{}
\begin{abstract}
Using initial-state radiation events from a total integrated
luminosity of 11.957 fb$^{-1}$ of $e^+e^-$ collision data collected at
center-of-mass energies between 3.773 and 4.258~GeV with the BESIII
detector at BEPCII, the cross section for the process $\EE\ar\llb$ is
measured in 16 $\llb$ invariant mass intervals from the production
threshold up to 3.00~GeV$/c^{2}$. 
The results are consistent with previous
results from BaBar and BESIII, but with better precision and with
narrower $\llb$ invariant mass intervals than BaBar.
\end{abstract}

\maketitle

\section{INTRODUCTION} \label{} Electromagnetic form factors (EMFFs),
which parametrize the inner structure of hadrons, are fundamental
observables for understanding the strong interaction. In the timelike
region, EMFFs are extensively studied in electron-positron
collisions by measuring hadron pair production cross sections. For a
spin-$1/2$ baryon ($B$), the cross section in the Born approximation
of the one-photon-exchange process $e^+e^-\to B\bar{B}$ is
parameterized in terms of electric and magnetic form factors $G_{E}$
and $G_{M}$ by~\cite{Bcs}:

\begin{equation}\label{Eq_Born1} \sigma^{B}(s) =
\frac{4\pi\alpha^{2}C\beta}{3s} \left[\left|G_{M}(s)\right|^{2} + \frac{2m^{2}_{B}c^{2}}{s}\left|G_{E}(s)\right|^{2}\right],
\end{equation}
where $\alpha$ is the fine-structure constant, $C$ is the Coulomb
correction factor~\cite{fcoulomb}, $\beta = \sqrt{1 -
  4m^{2}_{B}c^{4}/s}$ is a phase-space (PHSP) factor, $s$ is the
square of the center-of-mass (c.m.) energy, $m_{B}$ is the mass of the
baryon, and $c$ is the speed of light.  $C$ accounts for the
electromagnetic interaction of the fermions in the final state, and in
the point-like approximation, it is 1 for neutral baryons and
$y/(1-e^{-y})$ with $y = \pi\alpha\sqrt{1-\beta^{2}}/\beta$ for
charged baryons. Therefore, for charged baryon-pairs, the factor of
$\beta$ due to PHSP is canceled by the Coulomb factor, which results
in a non-zero cross section at the threshold when $\beta = 0$.
However, there is no cancelation in the neutral baryon-pair case, so
the cross section is zero.

There have been many experimental studies on the charged and neutral
baryon-pair production cross sections in the past decades, such as
$\EE\ar p\bar{p}$~\cite{ppbar, ppbarBaBar}, $\EE\ar
n\bar{n}$~\cite{nnbar}, $\EE\ar\Lambda\bar{\Lambda}$~\cite{BaBarcs,
  BEScs1, BEScs2, DM2cs}, $\EE\ar\Sigma\bar{\Sigma}$~\cite{Sig+-CS,
  Sig0CS}, $\EE\ar\Xi\bar{\Xi}$~\cite{X-X+CS, X0X0CS}, and
$\EE\ar\Lambda_{c}^{+}\bar{\Lambda}_{c}^{-}$~\cite{LLcCS}. Although
the conclusions for some channels are questionable due to large uncertainties, there is a
general tendency in the production cross sections for these baryon
pairs to have a step near the threshold, which then decreases with
the increase of the c.m.~energy of the baryon pair~\cite{BBbarCS}.

The cross section of the process $\EE\ar\llb$ very close to the
threshold has been measured in both the BaBar and the BESIII
experiments. In the BaBar experiment, the cross section from the
$\llb$ production threshold up to $M_{\llb}=$ 2.27~GeV$/c^{2}$ was
measured as $204_{-60}^{+62}\pm22$~pb~\cite{BaBarcs}. This result
indicates a possible non-zero cross section at threshold which is
in conflict with Eq.~(\ref{Eq_Born1}). However, due to the wide $\llb$
mass interval and large uncertainties, a solid conclusion cannot be
drawn. The BESIII experiment also measured the cross section at the
c.m.~energy ($\sqrt{s}$) of $2.2324$~GeV, which is only $1.0$~MeV above
the $\llb$ production threshold, to be
$305\pm45_{-36}^{+66}$~pb~\cite{BEScs1}. This indicates a threshold
enhancement phenomenon in the process $\EE\ar\llb$. Interestingly, in both 
the BaBar and BESIII experiments, a jump was observed in the process
$\EE\ar\kkkk$ near the $\llb$ production threshold~\cite{kkkkspe1,
  kkkkspe2}. 

To explain the near threshold enhancement, some theoretical studies
have been performed, in which the effects of final-state
radiation~\cite{FSR} and vector-meson resonances~\cite{vecreson1, vecreson2} have
been taken into account. The enhancement in the case of neutral
baryons may also be explained by an electromagnetic interaction
occurring at the quark level~\cite{quark}. However, experimentally,
the cross section measurements of $\EE\ar\llb$ near threshold are
still limited and more measurements are needed to further understand
this phenomenon.

\begin{figure}
	\includegraphics[width=2.67in]{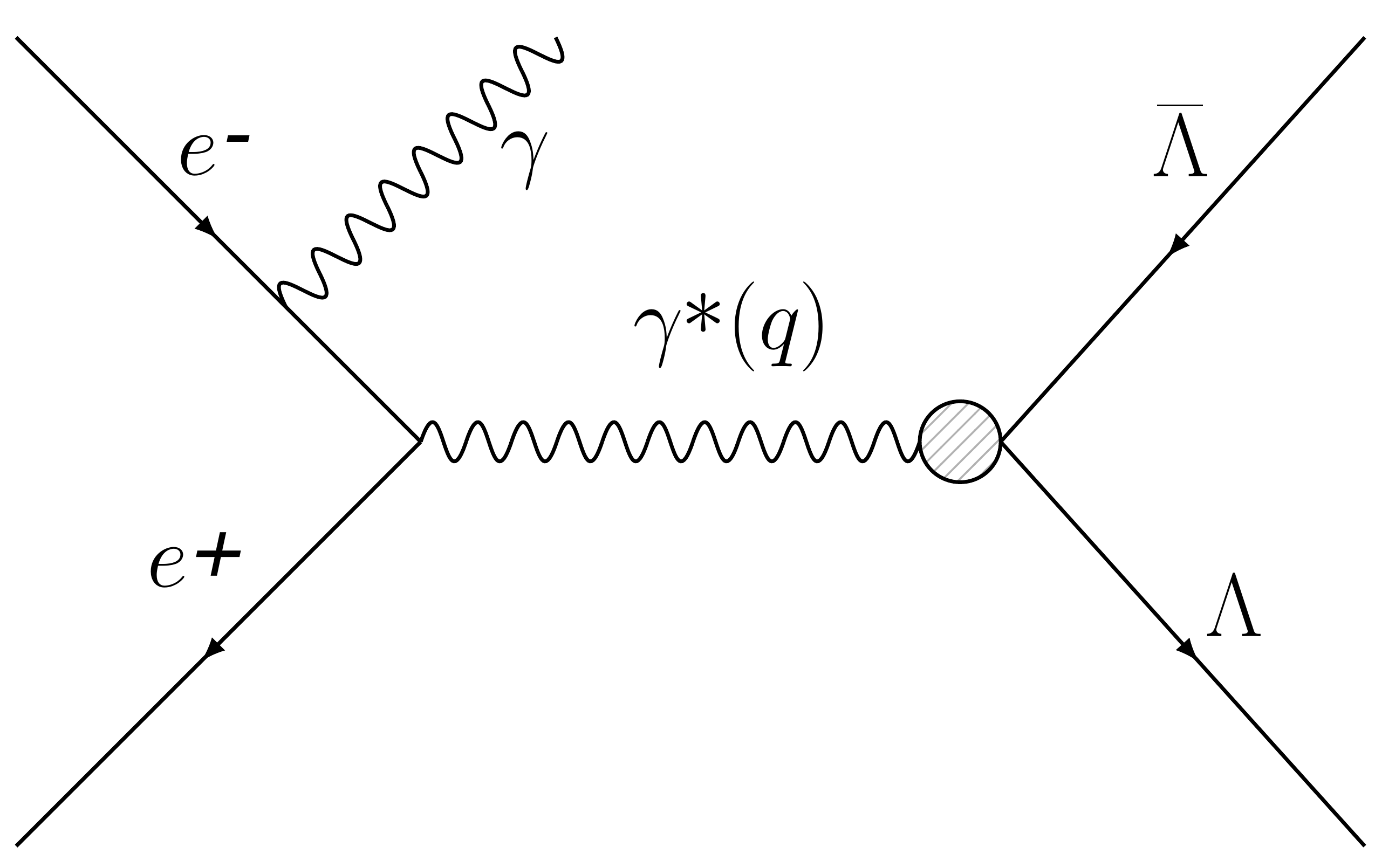}
	\caption{The leading-order Feynman diagram for the ISR process $\EE\ar\gamma\llb$. The ISR photon can be emitted from the electron or the positron.\label{feynman1}}
\end{figure}

The cross section and EMFFs of the $\Lambda$ hyperon have been
measured via the annihilation channel $\EE\ar\llb$ using the energy
scan technique~\cite{BEScs1, BEScs2, DM2cs}, in which the c.m. energy
of the collider is varied according to the experimental plan and the cross section is
measured at each c.m. energy. In addition, the radiative return channel
$\EE\ar\gamma\llb$ as illustrated in Fig.~\ref{feynman1}, where
$\gamma$ is a hard photon from the initial-state radiation (ISR)
process, offers a technique complementary to the energy scan technique
for the $\Lambda$ hyperon cross section measurement. This technique
has been used in the BaBar experiment to measure the cross section and
effective form factor of the $\Lambda$ hyperon~\cite{BaBarcs}.

The differential Born cross section for the $\EE\ar\gamma\llb$
process, integrated over the $\Lambda(\bar{\Lambda})$ momenta and the
photon polar angle, is written as~\cite{ISRfact2}:
\begin{equation} \label{ISRLamcs}
	\frac{d\sigma_{\EE\ar\gamma\llb}\left(q^2\right)}{dq^{2}}=\frac{1}{s}W(s, x)\sigma_{\llb}\left(q^{2}\right),
\end{equation} 
where $\sigma_{\llb}(q^2)$ is the cross section for the $\EE\ar\llb$ process, $q$ is the momentum transfer of the virtual photon whose squared value represents the invariant mass squared of $\llb$, $x=\frac{2E_{\gamma}^*}{\sqrt{s}}=1-\frac{q^2}{s}$, and $E_{\gamma}^*$ is the energy of the ISR photon in the $\EE$ c.m.~system.
The function~\cite{corr_ISRfact}
\begin{widetext}
	\begin{equation} \label{corr_ISRfact}
		\begin{aligned}
			W(s,x)=kx^{k-1}\left[1+\frac{\alpha}{\pi}\left(\frac{\pi^{2}}{3}-\frac{1}{2}\right)+\frac{3}{4}k+k^{2}\left(\frac{37}{96}-\frac{\pi^{2}}{12}-\frac{1}{72}\ln\frac{s}{m_{e}^{2}}\right)\right]-k\left(1-\frac{1}{2}x\right)\\
			+\frac{1}{8}k^{2}\left[4\left(2-x\right)\ln\frac{1}{x}-\frac{1+3\left(1-x\right)^{2}}{x}\ln\left(1-x\right)-6+x\right],	k=\frac{2\alpha}{\pi}\left[\ln\frac{s}{m_{e}^{2}}-1\right],
		\end{aligned}
	\end{equation}
\end{widetext}
describes the probability for the emission of an ISR photon with
energy fraction $x$, and $m_e$ is the electron mass.

In this analysis, we present the measurement of the $\EE\ar\llb$ cross
section from the production threshold up to 3.00~GeV$/c^2$ using the
ISR process $\EE\ar\gamma\llb$. The used data sets, corresponding to a
total integrated luminosity of 11.957 fb$^{-1}$, are collected at
twelve c.m.~energies between 3.773 and 4.258~GeV with the BESIII
detector~\cite{BESIII} at the BEPCII Collider ~\cite{BEPCII}.

\section{THE BESIII DETECTOR AND DATA SAMPLES}
The BESIII detector~\cite{BESIII} records symmetric $\EE$ collisions
provided by the BEPCII storage ring~\cite{BEPCII} in the
c.m.~energy range from 2.00 up to 4.95~GeV, with
a peak luminosity of $1\times10^{33}$ cm$^{-2}$s$^{-1}$ achieved at $\sqrt{s} = 3.77\;\text{GeV}$. BESIII has collected large
data samples in this energy region~\cite{BESIIIData}. The cylindrical
core of the BESIII detector covers 93\% of the full solid angle and
consists of a helium-based multilayer drift chamber (MDC), a plastic
scintillator time-of-flight system (TOF), and a CsI(Tl)
electromagnetic calorimeter (EMC), which are all enclosed in a
superconducting solenoidal magnet providing a 1.0~T magnetic
field~\cite{BESIII_Detec_new}. The solenoid is supported by an
octagonal flux-return yoke with resistive plate counter muon
identification modules interleaved with steel. The charged particle
momentum resolution at 1~GeV$/c$ is 0.5\%, and the d$E/$d$x$
resolution is 6\% for electrons from Bhabha scattering. The EMC
measures photon energies with a resolution of 2.5\% (5\%) at 1~GeV in
the barrel (end cap) region. The time resolution in the TOF barrel
region is 68~ps, while that in the end cap region used to be
110~ps. The end cap TOF system was updated in 2015 using multi-gap
resistive plate chamber technology, providing a time resolution of
60~ps~\cite{TOF1, TOF2, TOF3}.

The experimental data sets used in this analysis are listed in Table
\ref{datasamples}. To optimize the event selection criteria, Monte
Carlo (MC) simulations are performed with {\sc
  Geant4}-based~\cite{geant} software, which includes the description
of geometry and material, the detector response and the digitization
model, as well as a database for the detector running conditions and
performances. In this analysis, the event generator
\textsc{ConExc}~\cite{ConExc} is used to generate the signal process
$\EE\ar\gamma\llb$ ($\Lambda{\ar}p\pim$ $\bar{\Lambda}\ar\bar{p}\pip$)
with 1 million events at the different c.m.~energies up to ISR leading
order (LO), i.e. with only one ISR photon, and vacuum polarization
(VP) is included. The selection efficiencies are estimated by the
signal MC samples. An alternative event generator,
\textsc{PHOKHARA10.0}~\cite{phokhara}, is used to study the systematic
uncertainty of the MC model. The cross section lineshape used for the
generation of the signal MC samples is from
Ref.~\cite{Lambda_lineshape}. Inclusive MC samples at $\sqrt{s}=3.773$
and $4.178$~GeV are used to investigate possible background
contamination. They consist of inclusive hadronic processes ($\EE\ar
q\bar{q}$, $q=u, d, s$) modeled with the \textsc{LUARLW}~\cite{Lund}
at $\sqrt{s}=3.773$~GeV and \textsc{KKMC}~\cite{KKMC1, KKMC2} at
$\sqrt{s}=4.178$~GeV, and the ISR production of vector charmonium
states ($\EE\ar\gamma\jpsi$, $\gamma\psi(2S)$, $\gamma\psi(3773)$)
generated with \textsc{BesEvtGen}~\cite{eventgen1} using the
\textsc{VECTORISR} model~\cite{VECISR1, VECISR2}. In addition, several
exclusive MC samples are generated to study the background, with
different event generators and models.

\begin{table}
	\centering
	\caption{The c.m.~energy $\sqrt{s}$~\cite{EcmM1, EcmM2} and the integrated luminosity $\mathcal{L}_{\rm int}$~\cite{Lum3773, LumM2, LumM3} of the data sets used in the present analysis. \label{datasamples}}
	\begin{ruledtabular}
	\begin{tabular}{cc}
		$\sqrt{s}$ $($GeV$)$  &$\mathcal{L}_{\rm int}$ $($pb$^{-1})$\\
		\hline
		3.773             &2931.8  \\
		4.128             &401.5   \\
		4.157             &408.7   \\
		4.178             &3189.0  \\
		4.189             &526.7   \\
		4.199             &526.0   \\
		4.209             &517.1   \\
		4.219             &514.6   \\
		4.226             &1047.3  \\
		4.236             &530.3   \\
		4.244             &538.1   \\
		4.258             &825.7   \\
	\end{tabular}
    \end{ruledtabular}
\end{table}

\section{EVENT SELECTION \label{eventsel}} 
The complete process we study is
$\EE\ar\gamma\llb\ar\gamma(p\pi^-)(\bar{p}\pi^+)$, with the final
state $\gamma p\pi^-\bar{p}\pi^+$, where $\gamma$ is the ISR
photon. To provide a clean sample in the threshold region, the ISR
photon is detected (tagged). However, the differential cross section of the ISR
reaction (such as $\EE\ar\gamma\llb$) as a function of the ISR photon
polar angle reaches its highest value when the photon is emitted at a
small angle relative to the direction of the electron (or positron)
beam~\cite{ISRfact2}. Since this is out of the angular acceptance of
the EMC, photons falling in this region cannot be detected, resulting
in a reduction of signal efficiency. Moreover, the detection
efficiency is further reduced by the low momenta of the pions,
which, according to the study of the signal MC samples, are mostly
less than 0.2~GeV$/c$. We categorize the reconstruction of signal candidates into two
modes: mode I corresponds to fully reconstructed events, i.e. all
particles in the final state are identified; in mode II, a partial
reconstruction method with a missing pion is used to increase the
efficiency.

Charged tracks detected in the MDC are required to be within
$\left|\cos\theta\right|<0.93$, where $\theta$ is the polar angle with
respect to the $z$ axis, which is the symmetry axis of the MDC. The
distance of closest approach of each charged track to the
interaction point must be less than 30 cm along the $z$ direction and
less than 10 cm in the transverse plane. For each signal candidate, at
least three charged tracks are required.

The combined information of d$E/$d$x$ and TOF is used to calculate
particle identification (PID) probabilities for the pion, kaon, and
proton hypotheses, and the particle type with the
highest probability is assigned to the track.

\begin{figure}
	\includegraphics[width=3in]{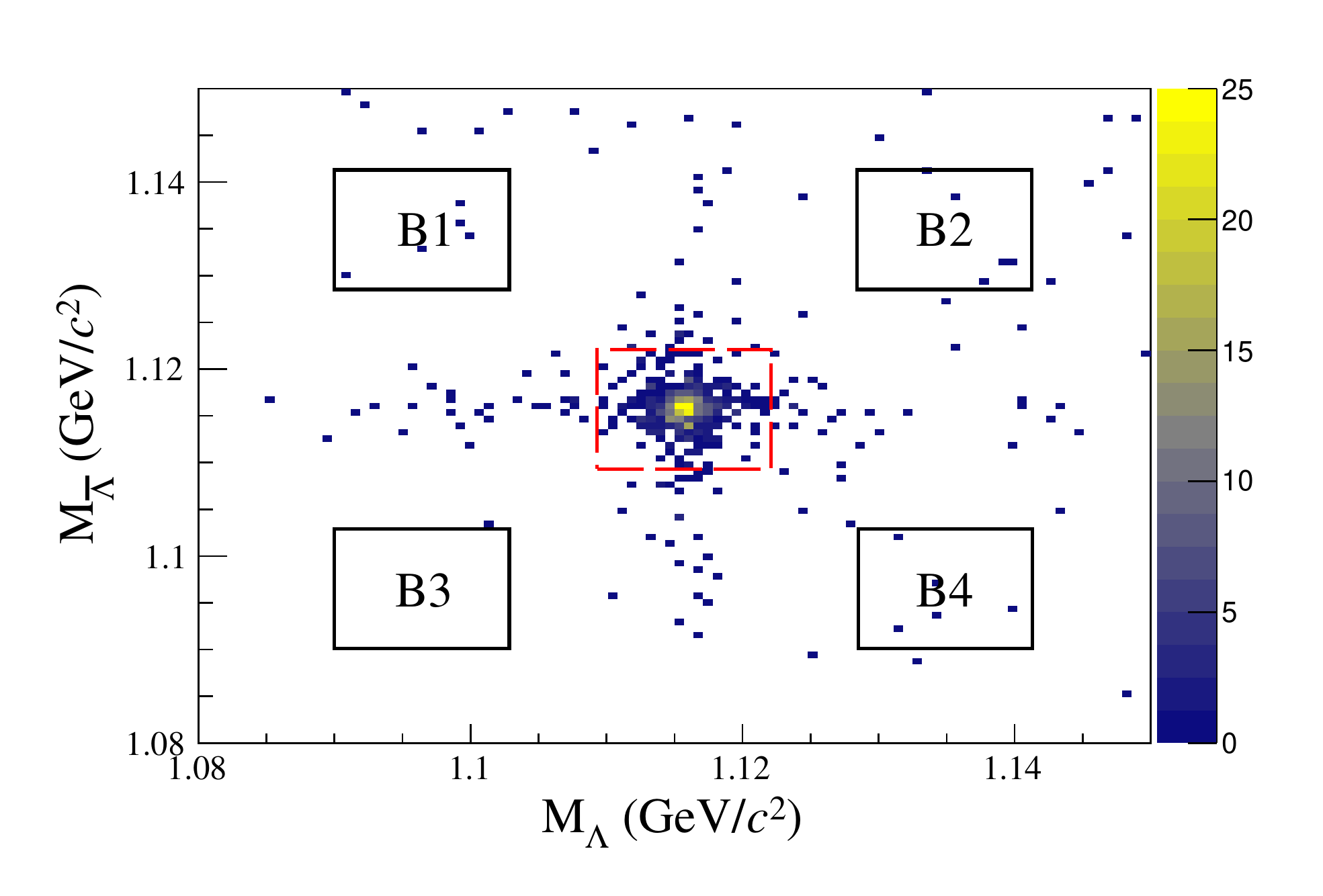}
	\caption{Distribution of $M_{\bar{\Lambda}}$ versus $M_{\Lambda}$ of the accepted candidates in mode I from all data sets. The dashed red box encloses the signal region, while the black boxes show the sideband regions. \label{2Dside}}
\end{figure}

A secondary vertex fit is performed to obtain the decay vertex of the
$\Lambda(\bar{\Lambda})$ candidate, and the $\Lambda(\bar{\Lambda})$
candidate is reconstructed by fitting the $p\pim(\bar{p}\pip)$ tracks
to a common decay vertex. If there is more than one
$\Lambda(\bar{\Lambda})$ candidate, the one with the minimum
chi-square value of the secondary vertex fit is selected. The reconstructed mass of
$\Lambda(\bar{\Lambda})$ candidate $\left(M_{\Lambda(\bar{\Lambda})}\right)$ is required to be within 6.4~MeV$/c^2$ of the nominal $\Lambda$ mass
($m_{\Lambda}$)~\cite{PDG}, as shown in
Fig.~\ref{2Dside}. There is no requirement on the decay length of
$\Lambda(\bar{\Lambda})$. Both a $\Lambda$ and a $\bar{\Lambda}$ are
required in mode I, while either a $\Lambda$ or a $\bar{\Lambda}$ is
required in mode II.

Information on the electromagnetic showers in the EMC is used to
select the photon candidates. It is required that the shower time is
within 700~ns of the event's start time to suppress electronic noise
and energy deposits unrelated to the events. A photon candidate is
selected if its deposited energy is greater than 0.4~GeV.
For each candidate signal event, at least one photon is required
which is considered as the ISR photon.

A kinematic fit is applied to further suppress background. For mode I, a
four-constraint (4C) kinematic fit requiring energy-momentum
conservation under the hypothesis of a $\gamma\llb$ final state is
applied to the signal candidates. If there is more than one
photon candidate, the combination with the minimum $\chi^2_{\rm 4C}$
is selected. To suppress the background with one more photon than the
signal process, we require $\chi^2_{\rm 4C}\leq\chi^2_{\rm
  4C,\gamma\gamma}$, where $\chi^2_{\rm 4C}$ and $\chi^2_{\rm
  4C,\gamma\gamma}$ are the chi-square values under the hypotheses of
$\gamma\llb$ and $\gamma\gamma\llb$ final states. For mode II, a one-constraint
(1C) kinematic fit with a missing $\pi^+(\pi^-)$ under the hypothesis
of a $\gamma\Lambda\bar{p}\pi^+(\gamma\bar{\Lambda}p\pi^-)$ final state
is applied to the signal candidates. Combining all
$\gamma\bar{p}(\gamma p)$ pairs with the reconstructed
$\Lambda(\bar{\Lambda})$, 1C kinematic fits are applied with the
invariant mass of $\bar{p}\pi^+(p\pi^-)$ being constrained to the
nominal $\Lambda$ mass~\cite{PDG} and the mass of $\pi^+(\pi^-)$ being unconstrained. The $\gamma\bar{p}(\gamma p)$
combination with the minimum $\chi^2_{\rm 1C}$ is selected, where
$\chi_{\rm 1C}^2$ is the chi-square of the 1C kinematic fit. A
requirement of $\chi^2_{\rm 4C}\leq50$ ($\chi^2_{\rm 1C}\leq5$) is
optimized for the signal candidates for mode I (mode II).

\begin{figure}
	\includegraphics[width=3in]{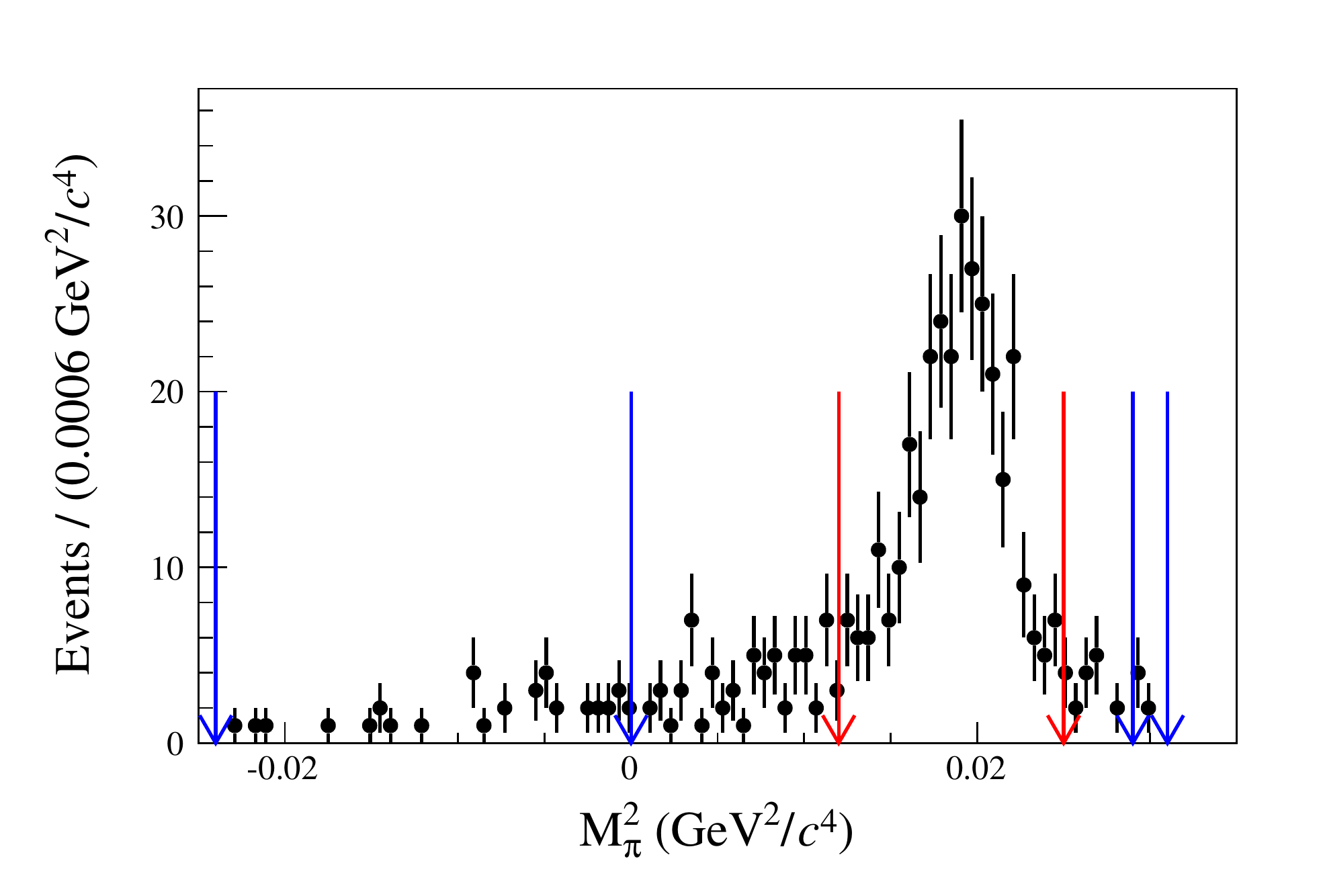}
	\caption{The $M_{\pi}^2$ spectrum of the accepted candidates
          in mode II from all data sets. The region between the red
          arrows is the signal region, and the regions between the blue
          arrows are the sideband regions. \label{1Dside}}
\end{figure}

For the candidates of mode II, the distribution of the mass squared
of the missing $\pi$ ($M_{\pi}^2$), obtained from energy-momentum
conservation, is shown in Fig.~\ref{1Dside}. To suppress background, a
requirement of $0.012\leq{M_{\pi}^{2}}\leq0.025$~GeV$^{2}/c^{4}$ is
applied.

\begin{figure}
	\includegraphics[width=3in]{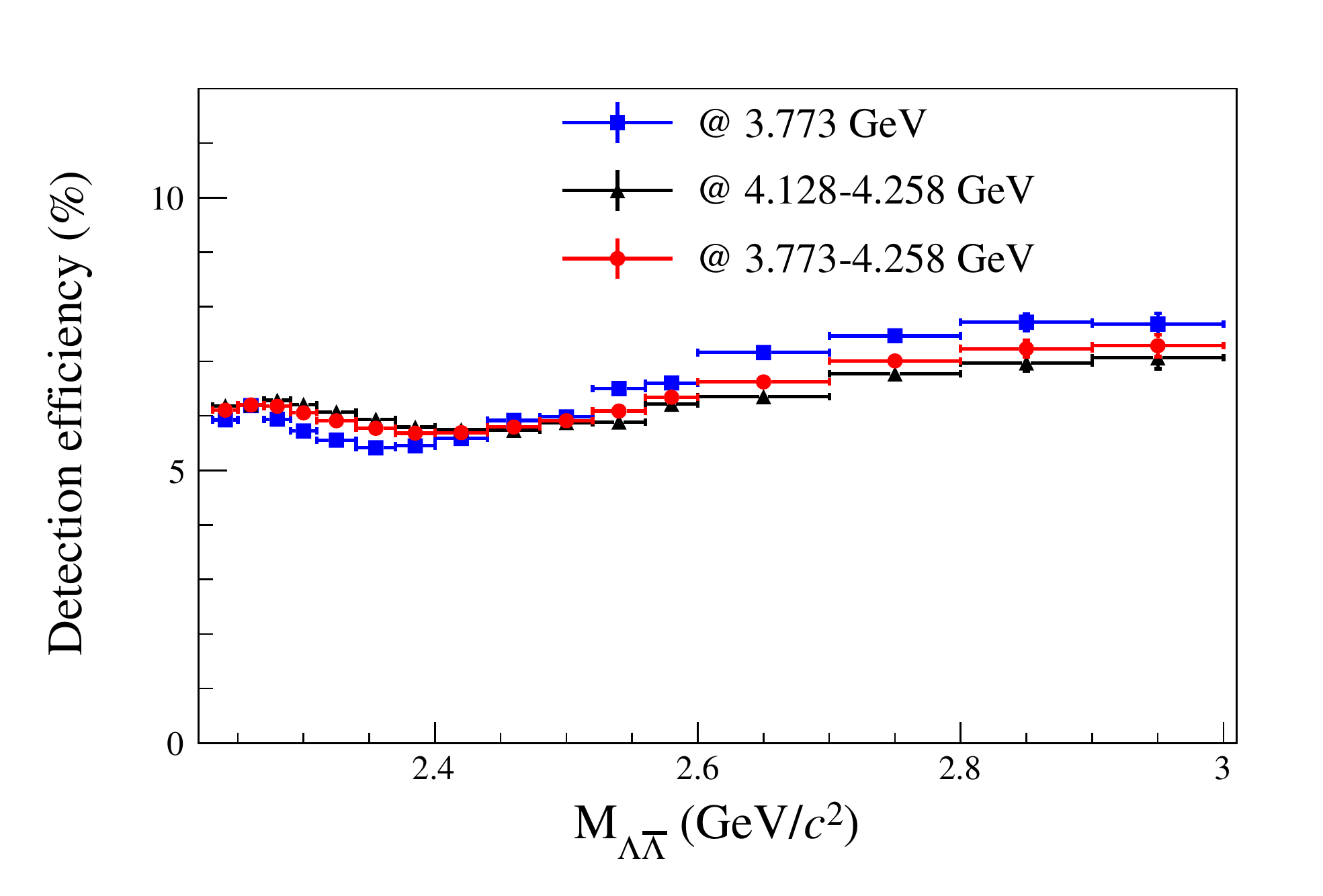}
	\caption{The $M_{\llb}$-dependent selection efficiencies
          obtained from MC simulation. Squares (blue), triangles
          (black), and circles (red) with error bars represent the
          data sets at $\sqrt s=$ 3.773, 4.178-4.258, and 3.773-4.258~GeV, respectively. The combined efficiency is weighted
          according to the effective luminosity of the ISR process.
          \label{effcurve}}
\end{figure}

\begin{figure}
	\includegraphics[width=3in]{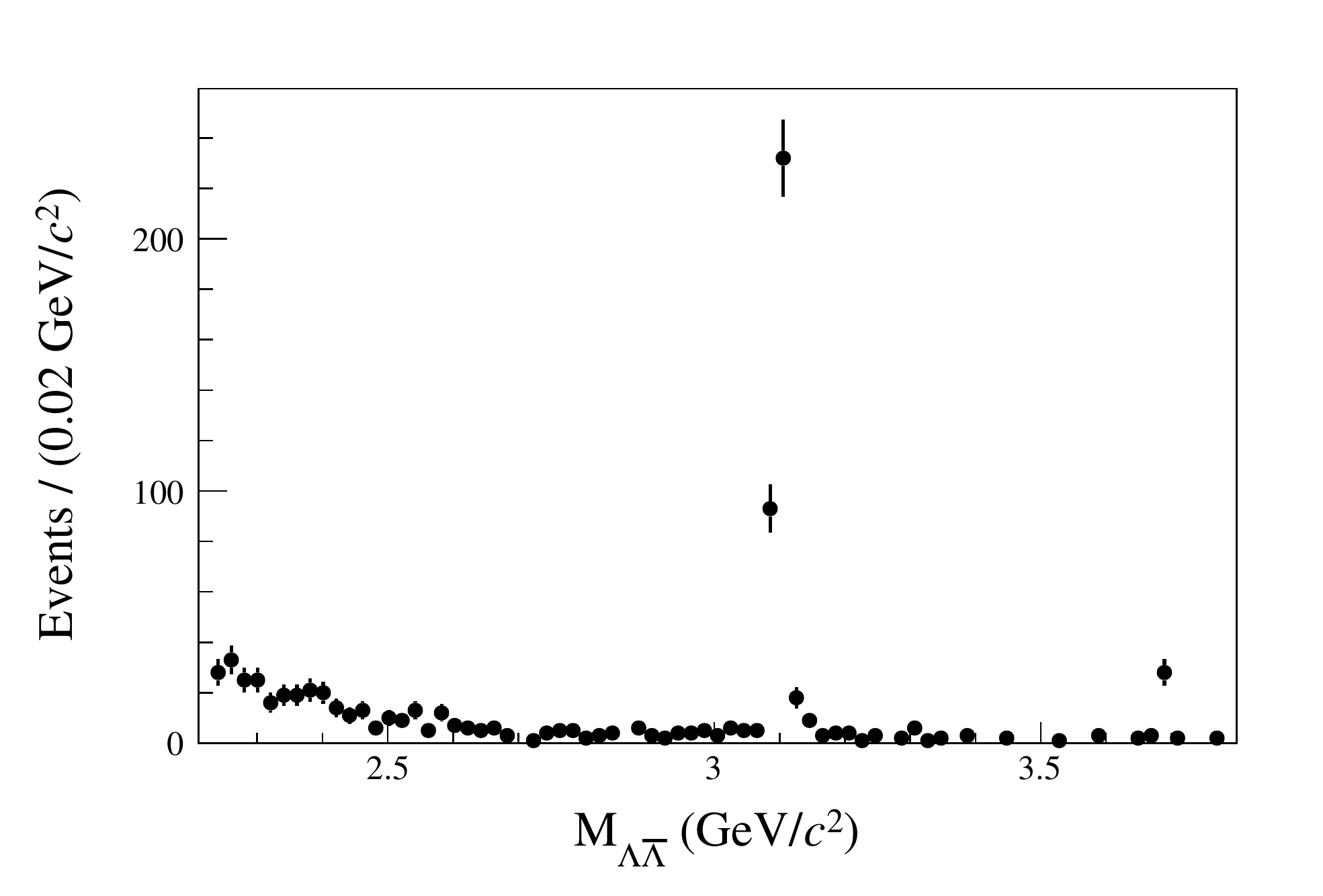}
	\caption{The $M_{\llb}$ spectrum for events satisfying the
          $\gamma\llb$ selection criteria from all data
          sets.  Contributions from $\jpsi\ar\llb$ and $\psi(2S)\ar\llb$
          decays are clearly seen.\label{Mlamlam}}
\end{figure}

The distribution of the selection efficiencies obtained from signal MC
samples as a function of invariant mass of $\llb$ ($M_{\llb}$) is
shown in Fig.~\ref{effcurve}, where the efficiencies at the
c.m.~energies between 4.128 and 4.258~GeV are combined and weighted
according to the effective luminosity of the ISR process. It should be noted that to
improve the mass resolution of $M_{\llb}$, we correct $M_{\llb}$ to
$\left(M_{\llb}-M_{\Lambda}-M_{\bar{\Lambda}}+2\times m_{\Lambda}\right)$. The mass resolution is given by the root-mean-square
deviation of $\left(M_{\llb}-M_{\llb}^{\rm truth}\right)$ of the
signal MC sample, where $M_{\llb}^{\rm truth}$ is the set value of the
invariant mass of $\llb$ when generating the MC events. In this paper,
the correction of the $M_{\llb}$ is implied unless specified. The
$M_{\llb}$ spectrum of the accepted candidates from all data sets is
shown in Fig.~\ref{Mlamlam}, in which 817 events are retained. The
contributions from $\jpsi\ar\llb$ and $\psi(2S)\ar\llb$ decays are
clearly seen.  About 60\% of the signal candidates have $M_{\llb}$
below 3.00~GeV/$c^2$, and the number of signal candidates ($N_{\rm
  obs}$) in each $M_{\llb}$ interval is listed in the first column of
Table~\ref{N_bkg_all}. 

\section{BACKGROUND ANALYSIS}
Potential background channels are investigated in the inclusive
MC samples with a topology analysis~\cite{topo}; they
consist of channels containing $\llb$ and channels without $\llb$. The
background channels containing $\llb$, such as the processes of
$\EE\ar\pi^0\llb$, $\EE\ar\gamma(\Lambda\bar{\Sigma}^0+c.c.)$, and
$\EE\ar\gamma\jpsi(\psi(2S))$ with $\jpsi(\psi(2S))$ decaying to
$\gamma\llb$, are studied individually, while the non-$\llb$ background is
estimated with the sideband method.

Events of $\EE\ar\pi^{0}\llb$ are easily mistaken as signal events
if a soft photon from the high-energy $\pi^{0}$ is missing. A
data-driven method is used to estimate their contribution. A sample
of $\pi^{0}\llb$ events is selected from data, and its background is
estimated with the sideband method. The sideband regions are chosen in
the distribution of the invariant mass of $\gamma\gamma$
($M_{\gamma\gamma}$). The number of events of this sample is
calculated by $N_{\pi^0}^{\rm data}=N_{\pi^0}^{\rm
  SigReg}-N_{\pi^0}^{\rm Side}/2$, where $N_{\pi^0}^{\rm SigReg}$ and
$N_{\pi^0}^{\rm Side}$ are the numbers of events from the signal and
the sideband regions of the $\pi^{0}\llb$ sample, respectively.  Next,
the contribution from the remaining $\pi^{0}\llb$ background
($N_{\pi^0}^{\rm bkg}$) in the signal candidates is determined by:
\begin{equation}
   	N_{\pi^0}^{\rm bkg}=N_{\pi^0}^{\rm data}\times\frac{N_{\rm ISR}^{\rm MC}}{N_{\pi^0}^{\rm MC}},
\end{equation}       
where $N_{\rm ISR}^{\rm MC}$ and $N_{\pi^0}^{\rm MC}$ are the numbers
of the events selected by the signal and $\pi^{0}\llb$ selection
criteria from the $\pi^{0}\llb$ MC sample. The $\pi^{0}\llb$ MC sample
is generated with the \textsc{ConExc}~\cite{ConExc} event generator up
to ISR LO, and the lineshape is obtained with the data sets collected at c.m. energies from 2.644~GeV to 3.080~GeV by BESIII.

In the reaction $\EE\ar\gamma(\Lambda\bar{\Sigma}^0+c.c.)$, the
$\Sigma^0(\bar{\Sigma}^0)$ decays to $\gamma\Lambda(\bar{\Lambda})$
with a branching ratio of 100\%~\cite{PDG}, where the $\gamma$ has low
energy. Therefore, if the photon from the $\Sigma^0(\bar{\Sigma}^0)$
decay is missing, this event can be misidentified as signal.  To
estimate the background from this reaction, a MC sample with a total
of 2 million events is generated with the
\textsc{ConExc}~\cite{ConExc} event generator up to ISR LO, and the
lineshape used to generate the MC events is determined with the data sets collected at c.m. energies from 2.309~GeV to 3.080~GeV by BESIII. After applying the signal ($\gamma\llb$)
selection criteria to this sample, we obtain the number of the
surviving $\gamma(\Lambda\bar{\Sigma}^0+c.c.)$ events
($N_{\Lambda\Sigma}^{\rm MC}$). A scaling factor is obtained by
$f=N_{\rm exp}/N_{\rm gen}$, where $N_{\rm exp}$ is the expected
number of the $\gamma(\Lambda\bar{\Sigma}^0+c.c.)$ events estimated
with the $(\Lambda\bar{\Sigma}^0+c.c.)$ cross section lineshape, and
$N_{\rm gen}$ is the number of MC simulated events. Finally, the
number of $\gamma(\Lambda\bar{\Sigma}^0+c.c.)$ background events
($N_{\Lambda\Sigma}^{\rm bkg}$) is estimated by
$N_{\Lambda\Sigma}^{\rm bkg}=f\times N_{\Lambda\Sigma}^{\rm MC}$.
Some other background channels, such as the processes $\EE\ar\eta\llb$
and $\EE\ar\gamma\jpsi(\psi(2S))$, are negligible.

Next, the sideband method is used to study the non-$\llb$
background. For mode I, two-dimensional (2D) sideband regions of
$M_{\Lambda}$ versus $M_{\bar{\Lambda}}$ are adopted, and for mode II,
one-dimensional (1D) sideband regions in the distribution of
$M_{\pi}^2$ are used. The distributions of
$M_{\Lambda(\bar{\Lambda})}$ and $M_{\pi}^2$ of inclusive MC samples
after removing the channels containing the $\llb$ pair are nearly
flat, so it is reasonable to use the sideband method. The 2D sideband
regions (shown in Fig.~\ref{2Dside}) are chosen as: B1: $1.0901\leq
M_{\Lambda}\leq1.1029$~GeV$/c^2$ and $1.1285\leq
M_{\bar{\Lambda}}\leq1.1413$~GeV$/c^2$; B2: $1.1285\leq
M_{\Lambda}\leq1.1413$~GeV$/c^2$ and $1.1285\leq
M_{\bar{\Lambda}}\leq1.1413$~GeV$/c^2$; B3: $1.0901\leq
M_{\Lambda}\leq1.1029$~GeV$/c^2$ and $1.0901\leq
M_{\bar{\Lambda}}\leq1.1029$~GeV$/c^2$; and B4: $1.1285\leq
M_{\Lambda}\leq1.1413$~GeV$/c^2$ and $1.0901\leq
M_{\bar{\Lambda}}\leq1.1029$~GeV$/c^2$.  The 1D sideband regions
(shown in Fig.~\ref{1Dside}) are chosen as $-0.024\leq
M_{\pi}^2\leq0$~GeV$^2/c^4$ and $0.029\leq
M_{\pi}^2\leq0.031$~GeV$^2/c^4$. The numbers of events from sideband
regions of data ($N_{\rm non-\llb}^{\rm data}$) are calculated by:
\begin{equation} \label{N_side}
	N_{\rm non-\llb}^{\rm data}=\frac{1}{4}\times N_{\rm 2D} + \frac{1}{2}\times N_{\rm 1D}, 
\end{equation} 
where $N_{\rm 2D}$ and $N_{\rm 1D}$ are the numbers of the events from the
2D and 1D sideband regions of data, respectively. The same sideband
regions are used for the $\pi^{0}\llb$ and
$\gamma(\Lambda\bar{\Sigma}^0+c.c.)$ MC samples, and the numbers of
events from sideband regions of these MC samples ($N_{\rm non-\llb}^{\rm MC}$) are
obtained with Eq.~(\ref{N_side}). The number of
non-$\llb$ background events ($N_{\rm non-\llb}^{\rm bkg}$) is
estimated by:
\begin{equation}
 	 N_{\rm non-\llb}^{\rm bkg}=N_{\rm non-\llb}^{\rm data}-N_{\rm non-\llb}^{\rm MC}.
\end{equation} 
The numbers of events for the three main background channels above ($N_{\pi^0}^{\rm bkg}$, $N_{\Lambda\Sigma}^{\rm bkg}$, $N_{\rm non-\llb}^{\rm bkg}$) are calculated in each $M_{\llb}$ interval when measuring the Born cross section.

\begin{figure}
	\includegraphics[width=3in]{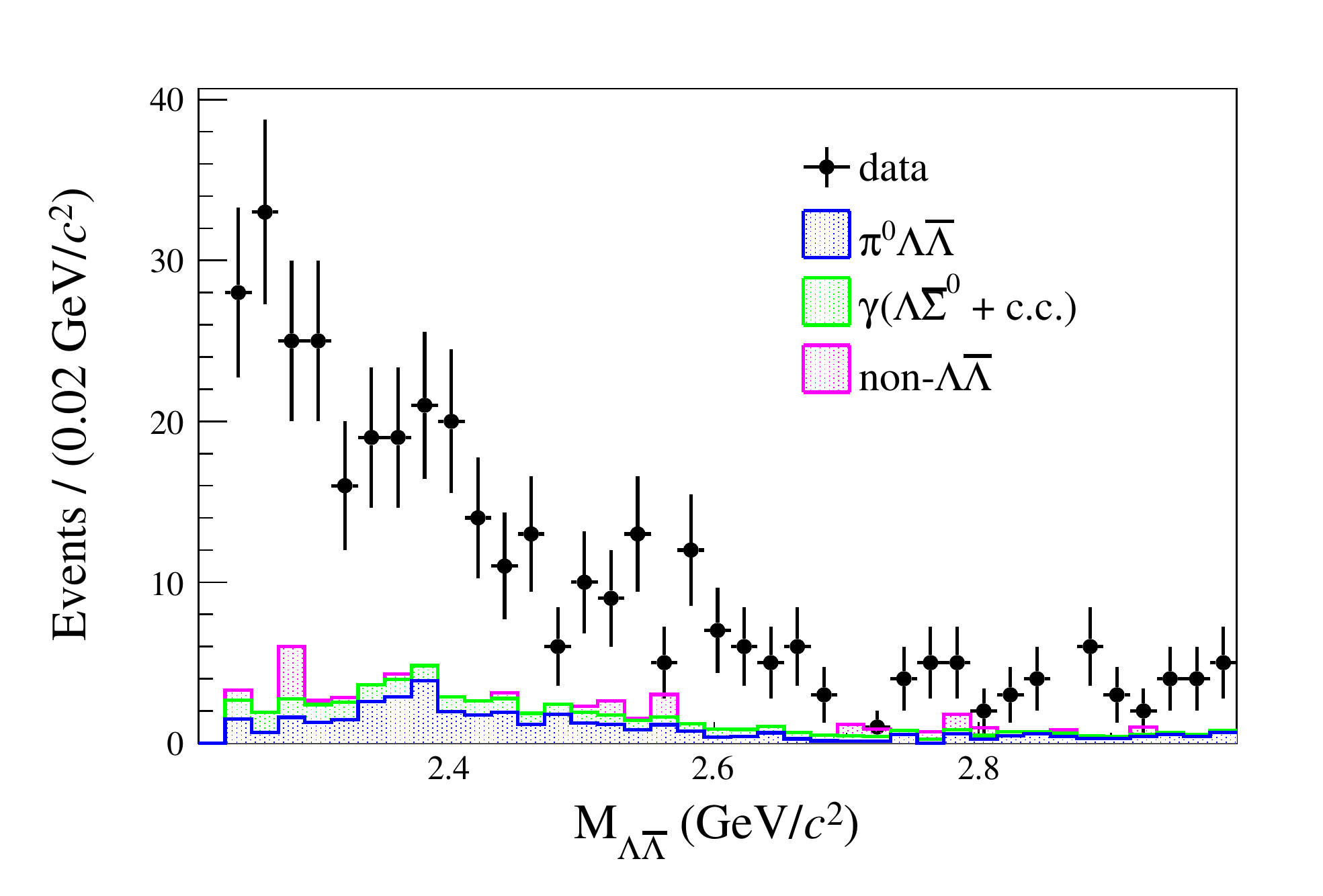}
	\caption{The distributions of $M_{\llb}$ for the signal
          candidates and the main background events from all data
          sets. Black dots with error bars refer to the signal
          candidates, and blue, green, and magenta histograms represent
          the $\pi^{0}\llb$, $\gamma(\Lambda\bar{\Sigma}^0+c.c.)$, and
          non-$\llb$ background events,
          respectively.  \label{Mlamlam_withbkg}}
\end{figure}

The distributions of $M_{\llb}$ of the main background events from all
data sets are shown in Fig.~\ref{Mlamlam_withbkg}, and the numbers of
background events over all data sets for the three main background
channels in each $M_{\llb}$ interval are listed in
Table~\ref{N_bkg_all}.
\begin{table}
	\centering
	\caption{The number of signal candidates ($N_{\rm obs}$),
          number of $\pi^{0}\llb$ events ($N_{\pi^0}^{\rm bkg}$),
          number of $\gamma(\Lambda\bar{\Sigma}^{0}+c.c.)$ events
          ($N_{\Lambda\Sigma}^{\rm bkg}$), and number of
          non-$\llb$ events ($N_{\rm non-\llb}^{\rm bkg}$), in each $M_{\llb}$
          interval, for the whole data set. The uncertainties are
          statistical. \label{N_bkg_all}}
	\begin{ruledtabular}
	\begin{tabular}{ccccc}
		\makecell[c]{$M_{\llb}$\\(GeV$/c^{2}$)}         &$N_{\rm obs}$     &$N_{\pi^0}^{\rm bkg}$    &$N_{\Lambda\Sigma}^{\rm bkg}$    &$N_{\rm non-\llb}^{\rm bkg}$    \\
		\hline
		2.231-2.250  &28.0 $\pm$ 5.3  &1.9 $\pm$ 1.2  &1.28 $\pm$ 0.05  &0.63 $\pm$ 0.70     \\
		2.25-2.27   &32.0 $\pm$ 5.7  &0.7$_{-0.5}^{+0.6}$     &1.35 $\pm$ 0.05  &$-$0.41$_{-0.02}^{+1.61}$ \\
		2.27-2.29   &25.0 $\pm$ 5.0  &1.4 $\pm$ 0.6     &1.36 $\pm$ 0.05  &2.67 $\pm$ 1.22  \\
		2.29-2.31   &24.0 $\pm$ 4.9  &1.3 $\pm$ 0.6     &1.37 $\pm$ 0.05  &0.69 $\pm$ 0.71     \\
		2.31-2.34   &28.0 $\pm$ 5.3 &2.4 $\pm$ 0.7     &2.00 $\pm$ 0.07  &0.08$_{-0.50}^{+1.24}$  \\
		2.34-2.37   &27.0 $\pm$ 5.2  &4.2 $\pm$ 0.9     &1.83 $\pm$ 0.05  &0.11$_{-0.50}^{+1.24}$  \\
		2.37-2.40   &34.0 $\pm$ 5.8  &5.2 $\pm$ 0.9     &1.54 $\pm$ 0.05  &$-$0.32$_{-0.02}^{+1.61}$  \\
		2.40-2.44   &28.0 $\pm$ 5.3  &3.5 $\pm$ 0.8     &1.74 $\pm$ 0.05  &0.10$_{-0.50}^{+1.24}$  \\
		2.44-2.48   &23.0 $\pm$ 4.8  &3.3 $\pm$ 0.7     &1.53 $\pm$ 0.05  &$-$0.32$_{-0.02}^{+1.61}$  \\
		2.48-2.52   &16.0 $\pm$ 4.0  &3.3 $\pm$ 0.7     &1.28 $\pm$ 0.05  &1.22$_{-0.87}^{+1.43}$  \\
		2.52-2.56   &19.0 $\pm$ 4.4  &1.7 $\pm$ 0.5     &1.01 $\pm$ 0.05  &1.51 $\pm$ 0.90    \\
		2.56-2.60   &18.0 $\pm$ 4.2  &1.4 $\pm$ 0.5     &0.87 $\pm$ 0.05  &$-$0.21$_{-0.02}^{+1.61}$   \\
		2.60-2.70   &24.0 $\pm$ 4.9  &1.4 $\pm$ 0.5     &1.74 $\pm$ 0.05  &$-$0.39$_{-0.02}^{+1.61}$  \\
		2.70-2.80   &15.0 $\pm$ 3.9  &1.5 $\pm$ 0.5     &1.12 $\pm$ 0.04  &3.00 $\pm$ 1.25     \\
		2.80-2.90   &15.0 $\pm$ 3.9  &2.3 $\pm$ 0.6     &0.73 $\pm$ 0.03  &0.07$_{-0.25}^{+1.17}$  \\
		2.90-3.00   &18.0 $\pm$ 4.2  &2.6 $\pm$ 0.7     &0.49 $\pm$ 0.03  &0.36$_{-0.50}^{+1.24}$    \\
	\end{tabular}
    \end{ruledtabular}
\end{table}

\section{SYSTEMATIC UNCERTAINTY}  
Several sources of systematic uncertainties are considered in the
cross section measurement. The combined results of different
reconstructed methods and different data sets are summarized in
Tables~\ref{corr_sys} and \ref{uncorr_sys} for the correlated and
uncorrelated parts, respectively. The correlated and uncorrelated
parts are summed in quadrature to determine the total uncertainty.
\begin{table}
	\centering
	\caption{The correlated systematic uncertainties (in \%) on
          the cross section measurement. $\mathcal{B}\left(\Lambda\ar
          p\pi\right)$ is the branching ratio of
          $\Lambda\left(\bar{\Lambda}\right)\ar
          p\pim\left(\bar{p}\pip\right)$.  \label{corr_sys}}
	\begin{ruledtabular}
		\begin{tabular}{lc}
			Source                                &Uncertainty\\
			\hline
			Luminosity                            &1.1 \\
			$\Lambda$ reconstruction              &2.1 \\
			$\bar{\Lambda}$ reconstruction        &2.8   \\
			$\mathcal{B}\left(\Lambda\ar p\pi\right)$ &1.6\\
			$p(\bar{p})$ tracking and PID           &0.7  \\
			$M^2_\pi$ window       &0.6   \\
			ISR photon detection                  &1.0   \\
			Kinematic fit                         &1.7   \\
			Neglected background                  &1.5   \\
			\hline
			Total                                 &4.7\\
		\end{tabular}
	\end{ruledtabular}
\end{table}
\begin{table}
	\centering
	\caption{The uncorrelated systematic uncertainties (in \%) in
          each $M_{\llb}$ interval on the cross section measurement:
          the uncertainty associated with the $\pi^0\llb$ channel
          ($\pi^0\llb$), $\gamma(\Lambda\bar{\Sigma}^0+c.c.)$ channel
          ($\gamma\Lambda\Sigma^0$), non-$\llb$ background
          (non-$\llb$), $\Lambda$ angular distribution (Ang), and
          signal MC model (MC). The last column is the total
          uncorrelated systematic uncertainty.  \label{uncorr_sys}}
	\begin{ruledtabular}
		\begin{tabular}{ccccccc}
			\makecell[c]{$M_{\llb}$\\(GeV$/c^{2}$)}    &$\pi^0\llb$ &$\gamma\Lambda\Sigma^0$ &non-$\llb$ &Ang  &MC                               &Total \\
			\hline
			2.231-2.250  &0.3   &0.6  &0.4   &2.7   &1.6     &3.2\\                             
			2.25-2.27   &0.1   &0.9  &1.4   &0.6   &1.4     &2.2\\
			2.27-2.29   &0.7   &1.8  &0.5   &2.3   &4.1     &5.1\\
			2.29-2.31   &0.9   &1.9  &0.4   &2.2   &0.7     &3.1\\
			2.31-2.34   &1.3   &3.6  &0.5   &2.7   &1.5     &4.9\\
			2.34-2.37   &0.8   &3.0  &0.4   &1.6   &0.9     &3.6\\
			2.37-2.40   &1.0   &2.0  &3.3   &0.3   &0.9     &4.1\\
			2.40-2.44   &0.6   &3.0  &0.4   &0.8   &0.8     &3.2\\
			2.44-2.48   &0.5   &2.4  &0.5   &1.7   &2.2     &3.6\\
			2.48-2.52   &1.6   &5.2  &5.2   &2.2   &2.2     &8.2\\
			2.52-2.56   &1.0   &5.4  &0.9   &1.7   &3.3     &6.7\\
			2.56-2.60   &0.5   &2.9  &0.4   &0.8   &1.8     &3.6\\
			2.60-2.70   &0.9   &3.2  &0.9   &2.5   &1.4     &4.4\\
			2.70-2.80   &7.1   &9.3  &24.8  &2.1   &1.9     &27.5\\
			2.80-2.90   &1.7   &2.3  &2.3   &2.1   &1.5     &4.4\\
			2.90-3.00   &1.2   &1.5  &0.3   &1.9   &5.4     &6.0\\
		\end{tabular}
	\end{ruledtabular}
\end{table}

The integrated luminosity is measured with an uncertainty of 0.5\% at
$\sqrt{s}=3.773$~GeV and an uncertainty of 1.0\% at other
c.m.~energies~\cite{Lum3773,LumM2,LumM3}. In this analysis, the
effective luminosity of the ISR process is calculated based on
Eq.~(\ref{corr_ISRfact}), and a 0.5\% uncertainty is
estimated~\cite{sysISR}. Thus, the total systematic uncertainty on the
luminosity is 0.8\% at $\sqrt{s}=3.773$~GeV and 1.2\% at other energy
points.

The uncertainties from the reconstruction of $\Lambda$ and
$\bar{\Lambda}$ are studied by a control sample of
$\jpsi\ar\ pK^-\bar{\Lambda}+c.c.$, and determined to be 2.8\% and
3.8\% at $\sqrt{s}=3.773$~GeV, and 2.6\% and 3.4\% at other energy
points, respectively.
A 1.0\% uncertainty is taken for the ISR photon
detection~\cite{photon_detec}.

For mode II, the uncertainties due to the $p(\bar{p})$ tracking and
PID are 1.0\% for each~\cite{proton_PID}. The uncertainty due to the
$M^2_{\pim}(M^2_{\pip})$ window is also studied by the control sample
of $\jpsi\ar\ pK^-\bar{\Lambda}+c.c.$, and estimated as 1.4\% (0.8\%)
at $\sqrt{s}=3.773$~GeV, and 1.5\% (0.9\%) at other energy points.
The uncertainty due to the branching fraction of
$\Lambda\left(\bar{\Lambda}\right)\ar p\pim\left(\bar{p}\pip\right)$,
$\mathcal{B}\left(\Lambda\ar p\pi\right)$, is obtained from the
PDG~\cite{PDG} to be 1.6\%.

The uncertainty from the kinematic fit is divided into two parts: the
contribution of the ISR photon and the contribution of the
remainder. The former is determined by a control sample of the radiative
Bhabha process $\EE\ar\gamma\EE$, and estimated as 0.4\%, 0.2\% and
1.1\% for the cases of full reconstruction, missing $\pim$ and missing
$\pip$, respectively. The later is studied by a control sample of
$\jpsi\ar\llb$, and is 0.2\% (0.2\%), 2.4\% (2.2\%) and
2.2\% (2.2\%) at $\sqrt{s}=3.773$~GeV (other energy points), for the
cases of full reconstruction, missing $\pim$ and missing $\pip$,
respectively. Thus, the uncertainty due to the kinematic fit at
$\sqrt{s}=3.773$~GeV (other energy points) is 0.6\% (0.6\%), 2.6\%
(2.4\%) and 3.3\% (3.3\%) for the cases of full reconstruction,
missing $\pim$ and missing $\pip$, respectively.

The signal MC samples are generated with PHSP. The angular distribution of the $\Lambda\bar{\Lambda}$ pair, the spin correlation between $\Lambda$ and $\bar{\Lambda}$, and the polarization of $\Lambda(\bar{\Lambda})$ decay are not taken into account. To estimate the uncertainty due to these factors, signal MC samples with an angular amplitude including these effects are generated. The parameterization of the angular amplitude is the same as that in Ref.~\cite{BEScs2}, and the corresponding parameters are cited from it when $M_{\Lambda\bar{\Lambda}}\leq2.52$~GeV$/c^{2}$ and obtained with the data set at $\sqrt{s}=2.900$~GeV when $M_{\Lambda\bar{\Lambda}}\geq2.52$~GeV$/c^{2}$.  The relative difference of the detection efficiency to
that of the PHSP mode is regarded as the uncertainty.

The uncertainty from the MC model is considered by changing the event
generator from \textsc{ConExc}~\cite{ConExc} to
\textsc{PHOKHARA10.0}~\cite{phokhara}. The relative difference of the
detection efficiency of these two event generators is taken as the
uncertainty.

For the channel of $\EE\ar\pi^0\llb$, the sideband regions on the
$M_{\gamma\gamma}$ spectrum are used to estimate the background of the
$\pi^0\llb$ sample. Here, the 2D sideband regions (sideband of
$M_{\Lambda}$ and $M_{\bar{\Lambda}}$) and 3D sideband regions
(sideband of $M_{\Lambda}$, $M_{\bar{\Lambda}}$ and
$M_{\gamma\gamma}$) are also used. The values of
$\left|\frac{N_{M_{\gamma\gamma}}-N_{\rm 2D}}{N_{\rm sig}}\right|$ and
$\left|\frac{N_{M_{\gamma\gamma}}-N_{\rm 3D}}{N_{\rm sig}}\right|$ are
obtained, where $N_{\rm sig}$ is the number of signal events,
$N_{M_{\gamma\gamma}}$, $N_{\rm 2D}$ and $N_{\rm 3D}$ are the
estimated numbers of $\pi^0\llb$ events based on $M_{\gamma\gamma}$,
2D and 3D sidebands, respectively. The larger of the two values is
taken as the uncertainty of this channel. 

For the channel of $\EE\ar\gamma(\Lambda\bar{\Sigma}^0+c.c.)$, one of
the parameters of the lineshape is changed by adding and subtracting a
standard deviation ($\pm1\sigma$). Based on the different lineshapes,
different estimated numbers of $\gamma(\Lambda\bar{\Sigma}^0+c.c.)$
events are obtained.  Further, the same method as for the $\EE\ar\pi^0\llb$
channel is used here to obtain the uncertainty of this channel.

For the non-$\llb$ background, we move the sideband regions 
by 0.002~GeV/$c^2$ and 0.002~GeV$^2/c^4$ towards the signal for
the 2D and the 1D sidebands, respectively, and
obtain the new estimated numbers of non-$\llb$ background events. The
relative difference between the old and new results is regarded as the
uncertainty. For the $M_{\llb}$ interval of 2.70-2.80~GeV$/c^2$, since
$N_{\rm sig}$ is extremely small ($0.8\pm2.3$) at
$\sqrt{s}=3.773$~GeV, the estimation of this uncertainty at
$\sqrt{s}=3.773$~GeV is significantly larger than that in other
intervals.
Except for the three main background sources mentioned above,
several other background channels are neglected, and their
contribution is considered as a systematic uncertainty, which is 2.2\%
at $\sqrt{s}=3.773$~GeV and 1.1\% at other energy points.

In this analysis, twelve data sets are used and three reconstruction methods (full reconstruction, and partial reconstruction with missing $\pim$ or $\pip$) are applied. We divide the data sets into two groups, where the first group only includes the data set at $\sqrt{s}=$ 3.773~GeV and the second group includes the other data sets at c.m.~energies from 4.128 to 4.258~GeV. The uncertainties of the second group are studied together or inherited from the result at $\sqrt{s}=4.178$~GeV. Thus, the systematic uncertainties are combined in two steps, where the first step combines the three reconstruction methods in each group and the second step combines the two groups. Uncertainties of the three reconstruction methods (two data set groups) are combined as the average value weighted by detection efficiencies (products of detection efficiency and effective luminosity). The weighted average formula is: 
\begin{equation} \label{ave_sys}
	\sigma_{\rm tot}^{2}=\sum_{i=1}^{3(2)}\omega_{i}^{2}\sigma_{i}^{2}+\sum_{i,j=1;i\neq j}^{3(2)}\rho_{ij}\omega_{i}\omega_{j}\sigma_{i}\sigma_{j},
\end{equation} 
with
\begin{equation} \label{weight}
	\omega_{i}=\frac{\varepsilon_{i}}{\sum_{i=1}^{3}\varepsilon_{i}}~\left(\omega_{i}=\frac{\varepsilon_{i}\mathcal{L}_{i}}{\sum_{i=1}^{2}\varepsilon_{i}\mathcal{L}_{i}}\right),
\end{equation}
where $\omega_{i}$, $\sigma_{i}$ and $\varepsilon_{i}$ with $i = 1, 2,
3$ ($i = 1, 2$) are the weight, systematic uncertainty and efficiency
for the reconstruction method (data set group) $i$, and $\rho_{ij}$ is
the correlation parameter for two different reconstruction methods
(data set groups) $i$ and $j$, and $\mathcal{L}_{i}$ is the effective
luminosity for the data set group $i$. For the systematic
uncertainties arising from background the $\rho_{ij}$ values are set
to 0, and for other systematic uncertainties the $\rho_{ij}$ are set
to 1.

\section{Results of the Cross section} \label{Sec7}
The cross section for $\EE\ar\llb$ is calculated from the $M_{\llb}$ spectrum by:
\begin{equation} \label{exp_CS}
	\sigma_{\llb}\left(M_{\llb}\right)=\frac{\left(dN_{\rm sig}/dM_{\llb}\right)}{\varepsilon\cdot \mathcal{B}^2\left(\Lambda\ar p\pi\right)\cdot d\mathcal{L}_{\rm int}/dM_{\llb}},
\end{equation} 
where $\left(dN_{\rm sig}/dM_{\llb}\right)$ is the $M_{\llb}$ spectrum
of data corrected for resolution effects after subtracting the
background, $\varepsilon$ is the detection efficiency from MC
simulation as a function of $M_{\llb}$, and $\mathcal{B}\left(\Lambda\ar p\pi\right) =
0.639 \pm 0.005$~\cite{PDG}. The effective ISR luminosity
$d\mathcal{L}_{\rm int}/dM_{\llb}$ is calculated by $d\mathcal{L}_{\rm
  int}/dM_{\llb}=W(s,x)\cdot\mathcal{L}_{\rm int}$, where $W(s,x)$ is
described by Eq.~(\ref{corr_ISRfact}). This effective luminosity
includes the first-order radiative correction but does not take into
account VP, so the obtained cross section is the ``dressed"
cross section.

\begin{figure}
	\includegraphics[width=3in]{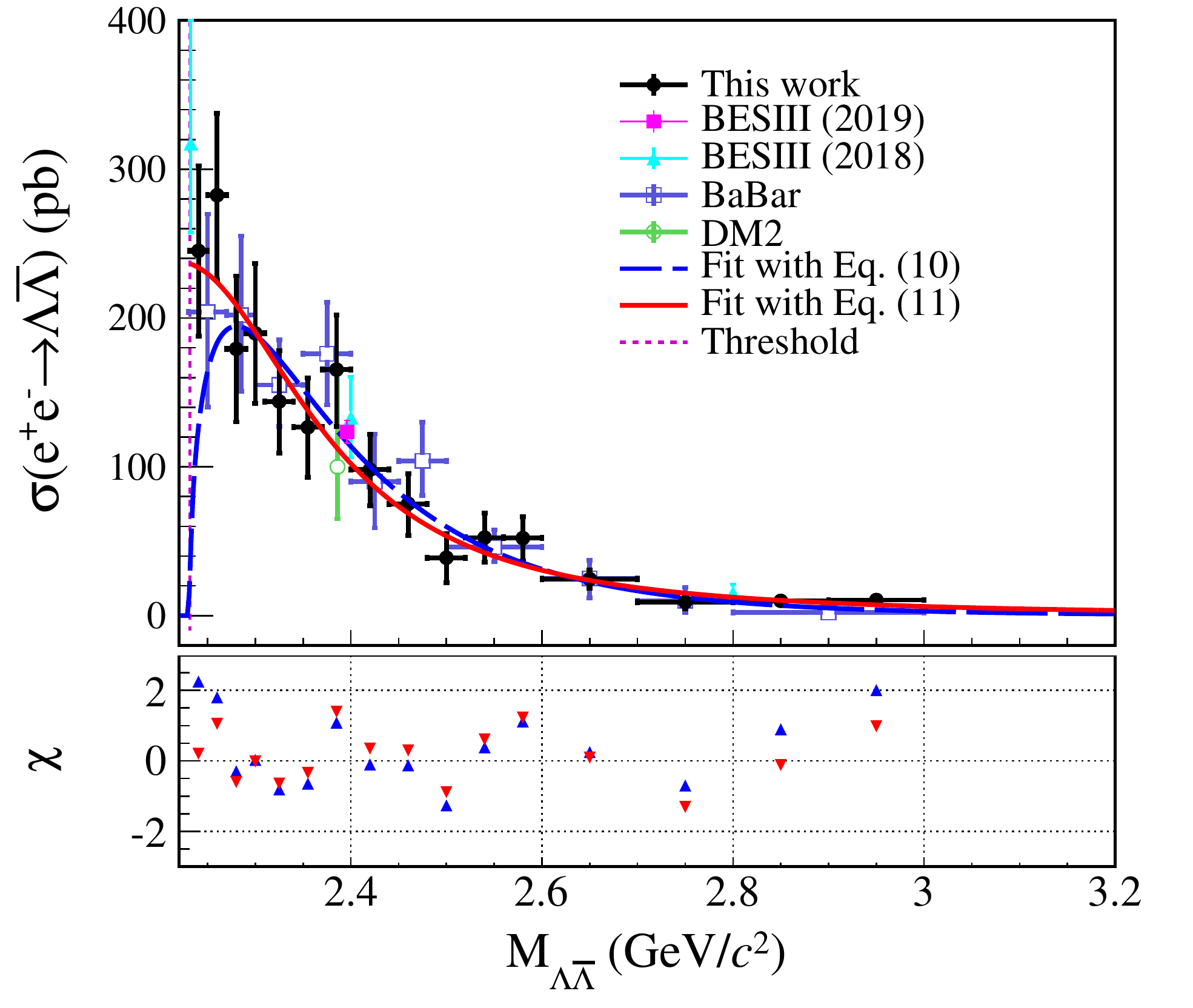}
	\caption{The cross section for the $\EE\ar\llb$ process from
          this analysis (black dots with error bars) with comparison
          to previous works (see the legend in the
          figure)~\cite{BEScs1, BEScs2, DM2cs, BaBarcs}. Both statistical and systematic uncertainties are included. The blue
          dashed line is the fit result using
          Eq.~(\ref{pQCD_func}), and the red solid line is the fit
          result using Eq.~(\ref{plateau}). The vertical dashed line
          is the production threshold for $\EE\ar\llb$. The
          $\chi$ distributions of the two fits are shown in the bottom
          panel, where the blue and red triangles represent the results of
          Eqs.~(\ref{pQCD_func}) and (\ref{plateau}), respectively.}
	\label{Fit_CS}	
\end{figure}

The dependence of the mass resolution on $M_{\llb}$ is determined, and
accordingly the $M_{\llb}$ is divided into 16 intervals from the
threshold up to 3.00~GeV$/c^2$. To reduce the impact of the mass
resolution, the width of the $M_{\llb}$ bin is at least 5 times
larger than the mass resolution, so we do not correct the mass spectrum for resolution effects. The measured cross sections for the process $\EE\ar\llb$ in these intervals are listed in
Table~\ref{CS_comb}. A comparison between the results of this work and
those of previous ones~\cite{BEScs1, BEScs2, DM2cs, BaBarcs} is
displayed in Fig.~\ref{Fit_CS}.
\begin{table}
	\renewcommand\arraystretch{1.15}
	\centering
	\caption{The cross section ($\sigma$) of the whole data
          set. $N_{\rm sig}$ is the total number of signal events,
          $\bar{\varepsilon}$ is the average detection efficiency of
          twelve energy points weighted by the effective ISR
          luminosity, and $\mathcal{L}$ is the total effective ISR
          luminosity. The uncertainties for $N_{\rm sig}$ are
          statistical. For $\sigma$, the first uncertainties are
          statistical, and the second are systematic.  \label{CS_comb}}
	\begin{ruledtabular}
		\begin{tabular}{ccccc}
			\makecell[c]{$M_{\llb}$\\ (GeV/$c^{2}$)}        &$N_{\rm sig}$     &$\bar{\varepsilon}$    &\makecell[c]{$\mathcal{L}$\\ (pb$^{-1}$)}    &\makecell[c]{$\sigma$ \\ (pb)} \\
			\hline
			2.231-2.250   &24.1 $\pm$ 5.5         &0.061   &3.95  &245 $\pm$ 56 $\pm$ 14\\
			2.25-2.27    &30.3$_{-5.9}^{+5.7}$  &0.062   &4.24  &283$_{-55}^{+53}$ $\pm$ 15\\
			2.27-2.29    &19.5 $\pm$ 5.2         &0.062   &4.32  &179 $\pm$ 48 $\pm$ 13\\
			2.29-2.31    &20.7 $\pm$ 5.0         &0.061   &4.41  &190 $\pm$ 46 $\pm$ 11\\
			2.31-2.34    &23.5$_{-5.5}^{+5.4}$  &0.059   &6.78  &144$_{-33.5}^{+32.7}$ $\pm$ 9.8\\
			2.34-2.37    &20.8$_{-5.4}^{+5.3}$  &0.058   &6.99  &126.6$_{-32.9}^{+32.1}$ $\pm$ 7.5\\
			2.37-2.40    &27.6$_{-6.1}^{+5.9}$  &0.057   &7.20  &165$_{-37}^{+35}$ $\pm$ 11\\
			2.40-2.44    &22.7$_{-5.5}^{+5.4}$  &0.057   &9.95  &98.1$_{-23.7}^{+23.2}$ $\pm$ 5.6\\
			2.44-2.48    &18.5$_{-5.1}^{+4.9}$  &0.058   &10.37 &75.2$_{-20.8}^{+19.7}$ $\pm$ 4.5\\
			2.48-2.52    &10.2$_{-4.3}^{+4.2}$   &0.059   &10.82 &38.9$_{-16.5}^{+15.9}$ $\pm$ 3.7\\
			2.52-2.56    &14.7 $\pm$ 4.5         &0.061   &11.30 &52.4 $\pm$ 16.0 $\pm$ 4.3\\
			2.56-2.60    &15.9$_{-4.6}^{+4.3}$  &0.063   &11.80 &52.1$_{-14.9}^{+14.0}$ $\pm$ 3.1\\
			2.60-2.70    &21.2$_{-5.2}^{+4.9}$  &0.066   &31.96 &24.6$_{-6.0}^{+5.7}$ $\pm$ 1.6\\
			2.70-2.80    &9.4 $\pm$ 4.1          &0.070   &35.96 &9.1 $\pm$ 4.0 $\pm$ 2.6\\
			2.80-2.90    &11.9$_{-4.1}^{+3.9}$  &0.072   &40.76 &9.9$_{-3.4}^{+3.3}$ $\pm$ 0.7\\
			2.90-3.00    &14.5$_{-4.5}^{+4.3}$  &0.073   &46.59 &10.5$_{-3.2}^{+3.1}$ $\pm$ 0.8\\		
		\end{tabular}
	\end{ruledtabular}
\end{table} 

A search for a threshold effect is made by performing a least chi-square fit to the cross section from the production threshold up to 3.00~GeV with different assumed functions. The systematic uncertainty is included in the fit with the correlated and uncorrelated parts considered separately. 
 
The first fit function is a perturbative QCD (pQCD) driven energy power function~\cite{pQCD}
\begin{equation}\label{pQCD_func}
	\sigma(s) = \frac{c_{0}\cdot \beta(s)\cdot C}{\left(\sqrt{s}-c_{1}\right)^{10}},
\end{equation}
where $c_0$ and $c_1$ are free parameters and the Coulomb correction
factor is $C=1$ for neutral baryons. The fit result is shown as the blue dashed line in
Fig.~\ref{Fit_CS}, with $c_0=(1.07\pm0.74)\times
10^3$~pb$\cdot$GeV$^{10}$, $c_1=1.27\pm0.08$~GeV and the fit quality
$\chi^2/d.o.f=19.06/14$.

In Fig.~\ref{Fit_CS}, the pQCD prediction does not describe
the anomalous enhancement well near threshold. Therefore, inspired
by the results of cross section measurements of $\EE\ar n\bar{n}$ and
$\EE\ar p\pbar$~\cite{nnbar, ppbar}, it is assumed that there is a
step near the threshold for the $\EE\ar\llb$ cross section,
the threshold enhancement effect. By taking into account the strong
interaction near the threshold instead of using the formula of
Eq.~(\ref{pQCD_func}), which contains the Coulomb factor, the cross
section can be expressed as~\cite{ppbar}:
\begin{equation}\label{plateau}
	\sigma(s) = \frac{e^{a_{0}}\pi^{2}\alpha^{3}}{s\left[1 - e^{-\frac{\pi\alpha_{s}}{\beta}}\right]\left[1 + \left(\frac{\sqrt{s} - 2m_{\Lambda}}{a_{1}}\right)^{a_{2}}\right]},
\end{equation}
where $a_{0}$, $a_{1}$, and $a_{2}$ are three free parameters. The
symbol $\alpha_s$ represents the strong running coupling constant and
is parameterized as:
\begin{equation}
	\alpha_{s} = \left[\frac{1}{\alpha_{s}(m_{Z}^{2})} + \frac{7}{4\pi}\ln\left(\frac{s}{m_{Z}^{2}}\right)\right]^{-1},
\end{equation}
where $m_{Z}=$ 91.1876~GeV$/c^{2}$~\cite{PDG} is the mass of $Z$ boson
and $\alpha_{s}(m_{Z}^{2})=$ 0.11856. This fit has
$\chi^{2}/d.o.f=9.83/13$, with $a_{0}=19.5\pm0.16$,
$a_{1}=0.17\pm0.04$~GeV and $a_{2}=1.98\pm0.34$, and the fit result is
shown as the red solid line in Fig.~\ref{Fit_CS}.

\section{\boldmath Study of the $\jpsi\ar\llb$ decay}
\begin{figure*}
	\includegraphics[width=6in]{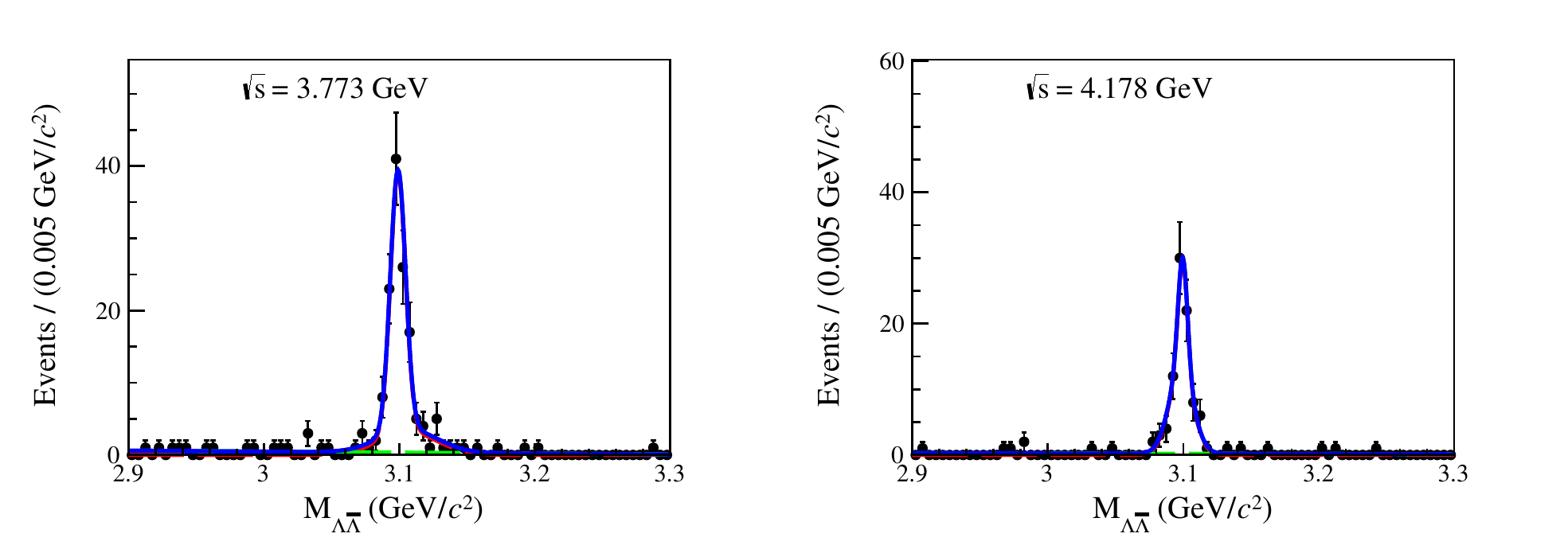}
	\caption{Simultaneous fit (blue curve) with a double Gaussian
		function (red dashed curve) for the resonance and a linear
		function (green dashed curve) for background of the
		$M_{\llb}$ spectra at $\sqrt{s}=3.773$ and
		$4.178$~GeV. Black dots with error bars represent
		data.  \label{BR_Jpsi}}
\end{figure*}

The branching fraction of $\jpsi\ar\llb$,
$\mathcal{B}\left(\jpsi\ar\llb\right)$, is determined via the ISR
process $\EE\ar\gamma\jpsi\ar\gamma\llb$ at $\sqrt{s}=3.773$ and
$4.178$~GeV. After integrating over the photon polar angle, the cross
section for ISR production of a narrow resonance (vector meson $V$),
such as $\jpsi$, decaying into the final state $f$ is given
by~\cite{CS_resonance}:
\begin{equation} \label{CS_resonance}
	\sigma(s)=\frac{12\pi^{2}\Gamma\left(V\ar\EE\right)\mathcal{B}(V\ar f)}{m_{V}s}W(s, x_{0}),
\end{equation}
where $m_{V}$ and $\Gamma(V\ar\EE)$ are the mass and electronic width
of the vector meson $V$, $x_{0}=1-m_{V}^{2}/s$, $\mathcal{B}(V\ar f)$
is the branching fraction of $V\ar\ f$, and $W(s, x_{0})$ is
calculated by Eq.~(\ref{corr_ISRfact}). If the cross section is
measured, the branching fraction can be calculated by
Eq.~(\ref{CS_resonance}). The cross section can also be written as:
\begin{equation}
	\sigma(s)=\frac{N_{\jpsi}}{\varepsilon\cdot\mathcal{B}^2\left(\Lambda\ar\ p\pi\right) \cdot\mathcal{L}_{\rm int}},
\end{equation} 
where $N_{\jpsi}$ is the number of $\jpsi$ events, $\varepsilon$ is
the detection efficiency, and $\mathcal{L}_{\rm int}$ is the
integrated luminosity of data, whose values are listed in
Table~\ref{datasamples}.  The detection efficiency is estimated from
MC simulation as 7.2\% at $\sqrt{s}=3.773$~GeV and 7.1\% at
$\sqrt{s}=4.178$~GeV. The angular distribution of $\Lambda$ in
$\jpsi\ar\llb$ decay is described by $1+\alpha\cos^2\theta_{\Lambda}$
with $\alpha=0.469$~\cite{LambAng}. To determine $N_{\jpsi}$, using
$\mathcal{B}\left(\jpsi\ar\llb\right)$ as a shared parameter, a
simultaneous fit is performed with a double Gaussian function for the
resonance and a linear function for the background and the continuum
contribution, and the  result is shown in Fig.~\ref{BR_Jpsi}

For the systematic uncertainties on the measurement of
$\mathcal{B}\left(\jpsi\ar\llb\right)$, the
uncertainties of the luminosity, $\Lambda$ and $\bar{\Lambda}$
reconstruction, $p(\bar{p})$ tracking and PID, $M^2_\pi$ window, ISR
photon detection, $\mathcal{B}\left(\Lambda\ar p\pi\right)$, and
kinematic fit are the same as the cross section measurement. The
uncertainty due to the MC model is assigned as 1.3\%, by changing the
model for the generation of the $\jpsi\ar\llb$ decay. The uncertainty
of the fit region is determined by changing the fit region from (2.90,
3.30)~GeV$/c^2$ to a wider (2.80, 3.30)~GeV$/c^2$ and a narrower
interval (3.00, 3.20)~GeV$/c^2$ to be 1.3\%. The uncertainty from the
signal model of the fit is estimated by changing the model from the double
Gaussian function to the MC-shape-convolved Gaussian function as
1.3\%. The uncertainty of the background model of the fit is estimated by
changing the model from a linear function to a constant as
0.5\%. Finally, we consider a systematic uncertainty due to the
non-$\llb$ background. The non-$\llb$ background is treated as a
peaking background, instead of a non-peaking one as default. The
relative difference between the results of the two strategies, 1.9\%,
is regarded as the uncertainty. The total uncertainty is obtained to
be 5.6\% by summing all uncertainties in quadrature.

$\mathcal{B}\left(\jpsi\ar\llb\right)$ is determined to be
$(1.64\pm0.12\pm0.09)\times10^{-3}$, where the first uncertainty is
statistical and the second is systematic. It is consistent with the
PDG value $(1.89\pm0.09)\times10^{-3}$~\cite{PDG} within 2$\sigma$.

\section{Summary and discussion}
Based on data sets corresponding to a total integrated luminosity of
11.957 fb$^{-1}$ collected at twelve c.m.~energies between 3.773 and
4.258~GeV with the BESIII detector at BEPCII, the cross section for
the process $\EE\ar\llb$ is measured as the function of $M_{\llb}$ in
16 intervals from the production threshold up to 3.00~GeV$/c^2$ using
ISR events with the ISR photon tagged. A partial reconstruction method
allowing a charged $\pi$ to be missing is used in addition to the full
reconstruction method to increase the efficiency. In the first
$M_{\llb}$ interval ranging from the threshold up to 2.25~GeV$/c^2$
(with the width of 19~MeV$/c^2$), the cross section is determined to
be $245\pm56\pm13$~pb, where the first uncertainty is statistical and
the second is systematic. It is a non-zero value with a statistical
significance of 4.3$\sigma$ and larger than the pQCD prediction by
2.3$\sigma$. In the region from 2.23~GeV$/c^2$ up to 3.00~GeV$/c^2$, the cross
section is measured in 15 intervals. The results are consistent with
previous measurements at BaBar and BESIII. The spectrum of the cross
section is fitted with the pQCD assumption and with the assumption of a
step existing near threshold, with the latter being a better description
of the data. 

\begin{acknowledgments}
The BESIII Collaboration thanks the staff of BEPCII, the IHEP computing center and the supercomputing
center of USTC for their strong support. This work is supported in part by National Key R\&D Program of China under Contracts Nos. 2020YFA0406400, 2020YFA0406300; National Natural Science Foundation of China (NSFC) under Contracts Nos. 11635010, 11735014, 11835012, 11935015, 11935016, 11935018, 11961141012, 12022510, 12025502, 12035009, 12035013, 12192260, 12192261, 12192262, 12192263, 12192264, 12192265, 12275320, 11625523, 11705192, 11950410506, 12061131003, 12105276, 12122509; the Chinese Academy of Sciences (CAS) Large-Scale Scientific Facility Program; the CAS Center for Excellence in Particle Physics (CCEPP); Joint Large-Scale Scientific Facility Funds of the NSFC and CAS under Contracts Nos. U1832207, U1732263, U1832103, U2032111; CAS Key Research Program of Frontier Sciences under Contracts Nos. QYZDJ-SSW-SLH003, QYZDJ-SSW-SLH040; 100 Talents Program of CAS; The Institute of Nuclear and Particle Physics (INPAC) and Shanghai Key Laboratory for Particle Physics and Cosmology; ERC under Contract No. 758462; European Union's Horizon 2020 research and innovation programme under Marie Sklodowska-Curie grant agreement under Contract No. 894790; German Research Foundation DFG under Contracts Nos. 443159800, 455635585, Collaborative Research Center CRC 1044, FOR5327, GRK 2149; Istituto Nazionale di Fisica Nucleare, Italy; Ministry of Development of Turkey under Contract No. DPT2006K-120470; National Research Foundation of Korea under Contract No. NRF-2022R1A2C1092335; National Science and Technology fund; National Science Research and Innovation Fund (NSRF) via the Program Management Unit for Human Resources \& Institutional Development, Research and Innovation under Contract No. B16F640076; Polish National Science Centre under Contract No. 2019/35/O/ST2/02907; The Royal Society, UK under Contracts Nos. DH140054, DH160214; The Swedish Research Council; U. S. Department of Energy under Contract No. DE-FG02-05ER41374.    
\end{acknowledgments}
	
\bibliography{main}
	
\end{document}